\documentclass[iopart,twocolumn,superscriptaddress,showpacs]{revtex4-1}
\usepackage{amssymb,amsmath,graphics,graphicx} 
\usepackage[usenames,dvipsnames,svgnames,table]{xcolor}
\usepackage{epstopdf}
\usepackage{enumerate}

\begin{document}
\title{Demonstrations of magnetic phenomena: Measuring the air permeablity using tablets}

\author{V.O.M. Lara}
\email[V.O.M. Lara: ]{vitor.lara@ifrj.edu.br}
\address{Instituto de F\'isica - Universidade Federal Fluminense, Niter\'oi - Rio de Janeiro Brazil}
\address{Instituto Federal de Educa\c{c}\~ao, Ci\^encia e Tecnologia do Rio de Janeiro, S\~ao Gon\c{c}alo - Rio de Janeiro Brazil}
\author{D. F. Amaral}
\address{Cons\'orcio de Ensino \`a dist\^ancia do Rio de Janeiro (CEDERJ), S\~ao Gon\c{c}alo - Rio de Janeiro Brazil}
\author{D. Faria} 
\address{Instituto de F\'isica - Universidade Federal Fluminense, Niter\'oi - Rio de Janeiro Brazil}
\author{L. P. Vieira}
\address{Instituto de F\'isica - Universidade Federal do Rio de Janeiro, Rio de Janeiro - Rio de Janeiro Brazil}

\pacs{ }

\date{\today}

\begin{abstract}
\textbf{Abstract} \\
We use a tablet to determine experimentally the dependencies of the magnetic field (B) on the electrical current and on the axial distance from a coil (z). Our data shows a good precision on the inverse cubic dependence of the magnetic field on the axial distance, $B \propto z^{-3}$. We obtain with good accuracy the value of air permeability $\mu_{air}$. We also observe the same dependence of $B$ on $z$ when considering a magnet instead of a coil. Although our estimates are obtained through simple data fits, we also perform a more sophisticated error analysis, confirming the result for $\mu_{air}$.

\end{abstract}

\maketitle


The use of tablets and smartphone in science education expands possibilities for approaches that motivate students to understand better several physical phenomena\cite{previous_mag,artigo_gota, artigo_ondas, diss_Leo}. 
In particular, tablets were shown as good tools to measure magnetostatic responses in current-carrying wires. This interesting work is about magnetic field sensoring \cite{previous_mag}. It reports a simple way of obtaining experimentally the linear dependence between the magnetic field $B$ and the number of turns $N$ in the current-carrying coil using an "app" for iPad \cite{magnetometer_app}. However, additional dependences of B are still not discussed, some of which we show in this paper leading to a wider description of this kind of system.
 
We determine the dependencies of $B$ on the electric current $I$ and on the axial distance $z$ in a coil, in suitably conditions using an iPad and the same free app MagnetMeter \cite{magnetometer_app} (we suggest a similar app for Android \cite{android_app}). We also perform the similar experiments with a small magnet instead of a coil. For the coil, we also make a good estimate for the magnetic permeability of the air $\mu_{air} \cong \mu_{0} \equiv 4\pi \times 10^{-6} Hm^{-1}$. 

\begin{figure}[h!]
\begin{center}
\includegraphics[scale=0.21]{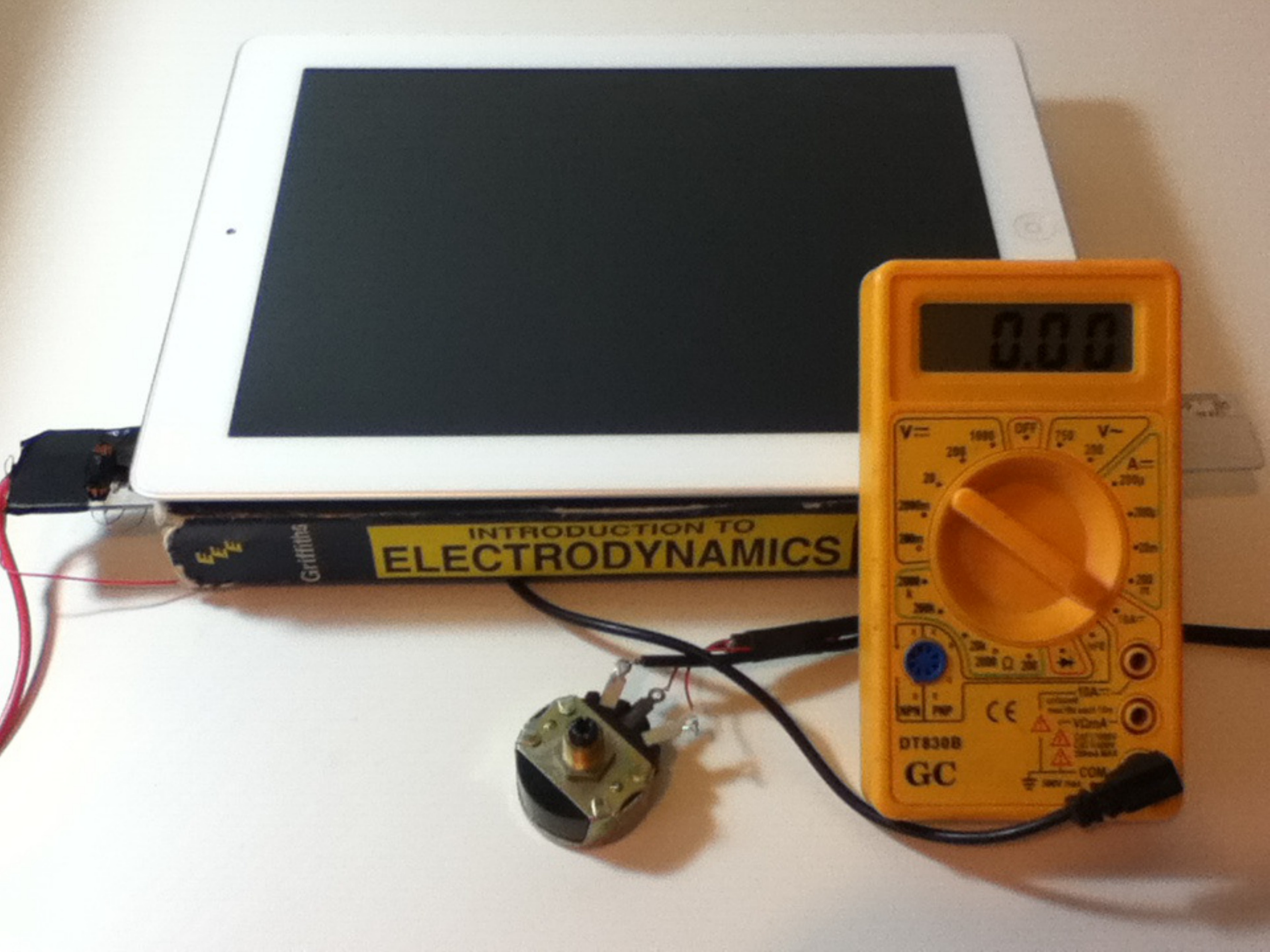}
\end{center}
\caption{(Color online) Photography from the experimental setup.} 
\label{fig:demonstration_set}
\end{figure}

The demonstration set used is composed by an electrical circuit, a ruler and a book. The circuit is formed by the following components, all of them connected in series: a wirewound potentiometer with resistance up to $30\Omega$; a resistor with $10\Omega$; an electrical source from a cell phone (max. output current $\sim0.9A$); a digital multimeter and a coil (internal diameter $2R_{i} = 1.91 \; cm$ and external diameter $2R_e = 2.42 \; cm$ and $N = 62$ turns). 

\begin{figure}[h!]
\begin{center}
\includegraphics[scale=0.35]{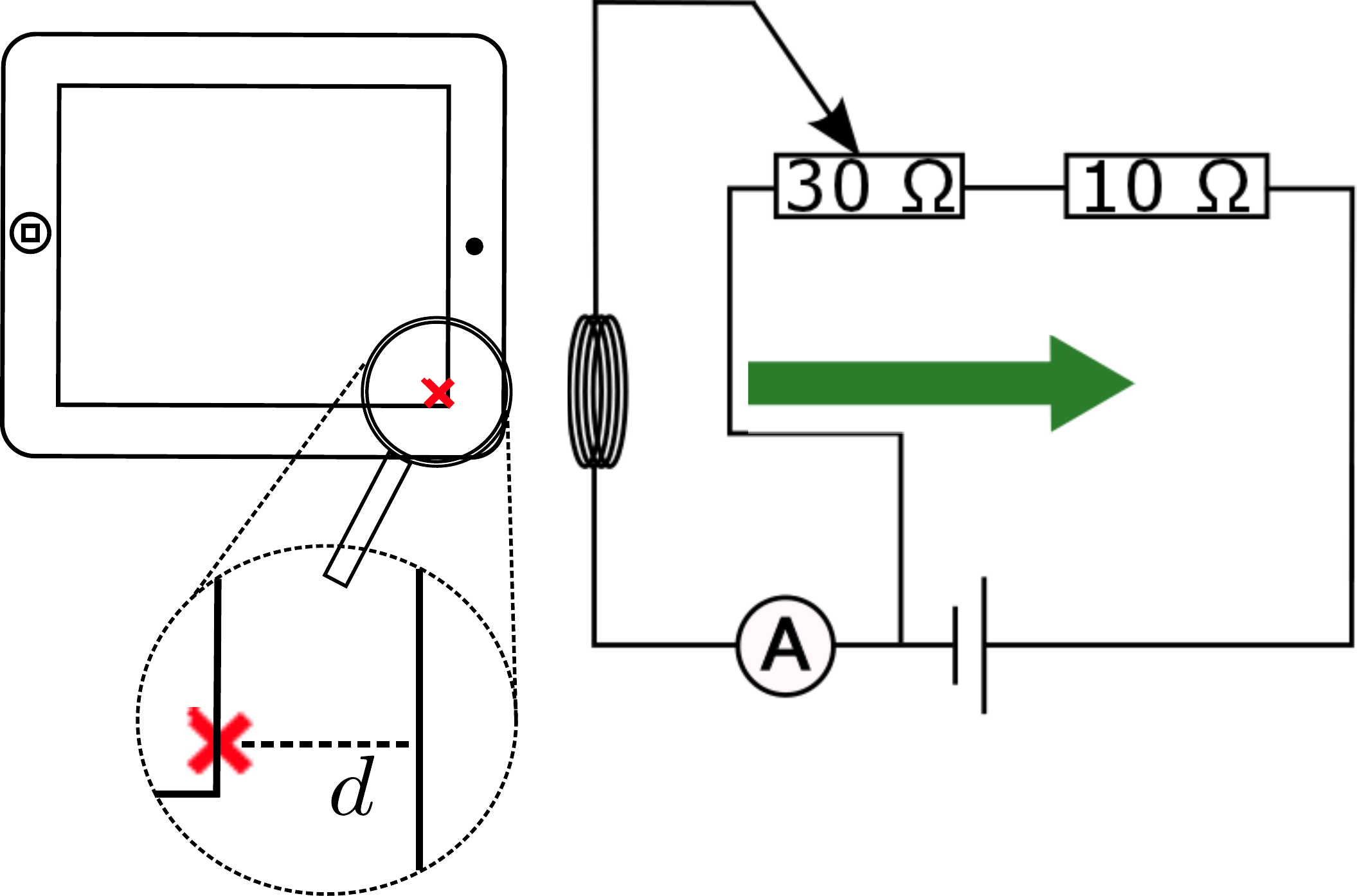}
\end{center}
\caption{Schematic representation of the coil experiment.  The red cross indicates the magnetometer position inside the iPad, while the green arrow corresponds to the z axial displacement of the coil.} 
\label{fig:mesh_3}
\end{figure}

Next, we describe the circuit assembly, which is relatively easy to built. This circuit is formed in such a way that the potentiometer enables the variation of the current in the coil, which is measured by the ammeter. However, for safety issues, it could also be necessary to add the extra resistance of $10\Omega$ to avoid high currents. The potentiometer has three terminals. The middle one has to be connected to the coil, while each one of the other terminals are connected to the resistor and to the negative source terminal, as it can be seen in figure \ref{fig:mesh_3}. There is no need to worry about the connection order of these two terminals, because the circuit should behave as expected in both ways. Although, one should be careful about the direction in which the potentiometer will increase the current value measured by the ammeter. 

\begin{figure}[h!] 
\centering\includegraphics[scale=0.67]{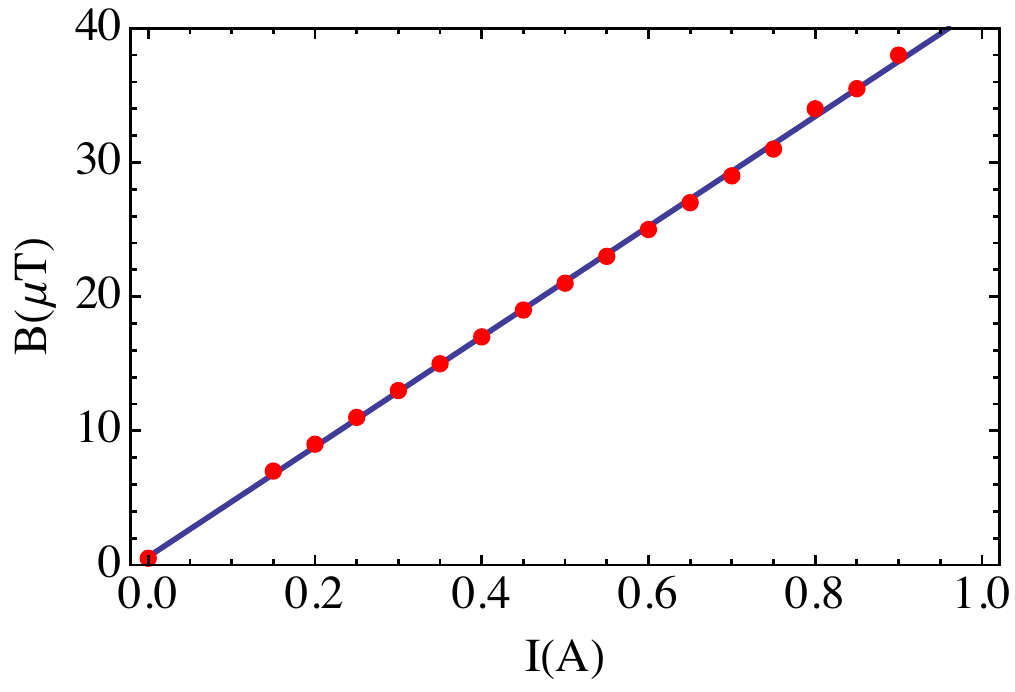}
\caption{(Color online) Magnetic field linear dependence on the electric current for the coil. The blue line represents the linear fit and the red points correspond to the obtained data. The values of $a$ and $b$ vary according to the coil radii and to the axial distance from the magnetic sensor.  } 
\label{fig:BXi} 
\end{figure}

As a standard procedure throughout, before starting the experiment we press the red button on the Magnetmeter app in order to set any other relevant magnetic interferences aside, such as the Earth's magnetic field. 
In the first experiment we pulled the coil up close to the iPad upper right edge (see figure \ref{fig:mesh_3}). We fixed the axial distance between the coil and the magnetometer at $z=4.8\;cm$. It is crucial that one takes into account the distance $d$ relative to the localization of the magnetic sensor inside the iPad, adding it to the value measured by the ruler (for the iPad we use $d \sim 1.8 \; cm$\cite{footnote}). Next we increase the current by equal amounts $\delta I = 0.05 \; A$, writing down the magnetic fields measured by each correspondent current, as plotted in figure \ref{fig:BXi}. 
The data adjust was realized using the`fit' command from the \textit{Gnuplot}\cite{gnuplot}, with 

\begin{eqnarray}
 B(I) &=& aI + b,     \nonumber \\
   a &=& (41.0247\pm 0.2571 ) \mu T/A   ~~\mbox{and} \nonumber \\
   b &=& (0.611347 \pm 0.143 )\mu T\,\,\,.
\end{eqnarray}

\begin{figure}[!htb]
\begin{center}
\includegraphics[scale=0.65]{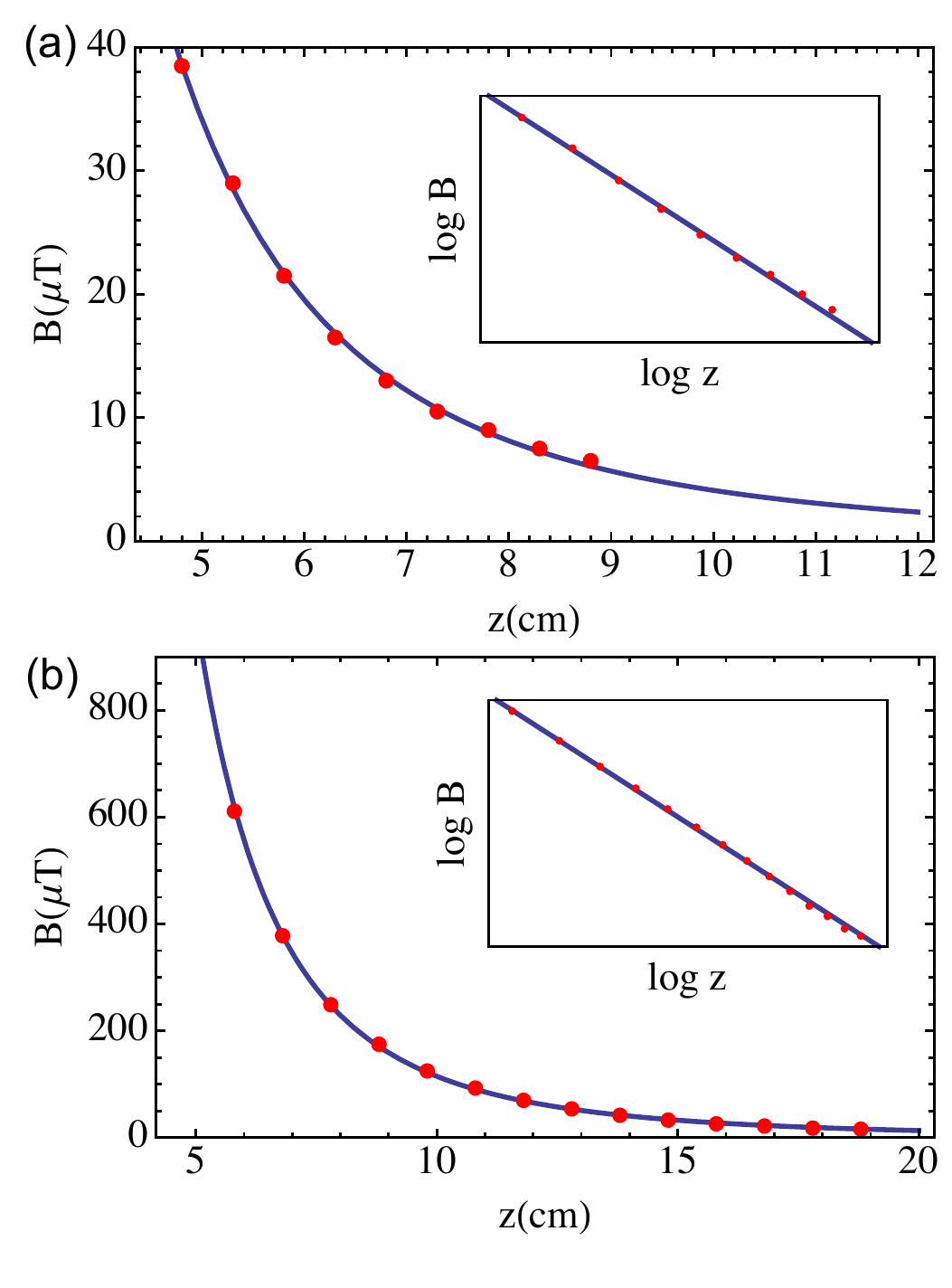}
\end{center}
\caption{(Color online) Magnetic field versus distance for the (a) coil and (b) magnet. The inset shows the same data on $\log \log$ scale.}
\label{fig:B_X_d_coil}
\end{figure}

The second experiment consists of an analysis of the magnetic field dependence on $z$ for both the coil and the magnet. We start by holding the ruler between the book pages and positioning the iPad above the book with its magnetic sensor facing the coil or the magnet. For the coil we increase the current up to its maximum $I\sim 0.9 \; A$, which is not necessary in the case of the magnet due to its permanent magnetization.  We subsequently move the coil (magnet) by equal displacements in the green arrow direction as indicated in figure \ref{fig:mesh_3}. We take notes of the magnetic field showed by the app for each distance. In figure \ref{fig:B_X_d_coil} we plot the experimental data of $B$ as a function of $z$ obtained by the Demonstration Set for (a) the coil (see Table \ref{table:tabela2}) and (b) the magnet, respectively.
In addition, we perform a data fit using $B\left(z\right) = a z^{b}$, as shown in table \ref{table:tabela1}. For both cases we obtain an excellent agreement with the expected $z^{-3}$ dependence \cite{griffiths}.

\begin{table}[h!]
\caption{ Magnetic field fitting, $B\left(z\right) = a z^{b}$ for the coil and for the magnet.}
\begin{tabular}{ | l | c | c | p{5cm} |}
\hline
    $Parameters$ & $a \pm \delta a$  & $b \pm \delta b$ \\ \hline
    $Coil$  & $4624.6 \pm 307.4 $ & $-3.05112 \pm 0.03917$ \\ \hline
    $Magnet$ & $142012 \pm 2903$ & $-3.09232 \pm 0.01217$ \\ \hline
\end{tabular}
\label{table:tabela1}
\end{table}

\begin{figure}[!htb]
\includegraphics[scale=0.1]{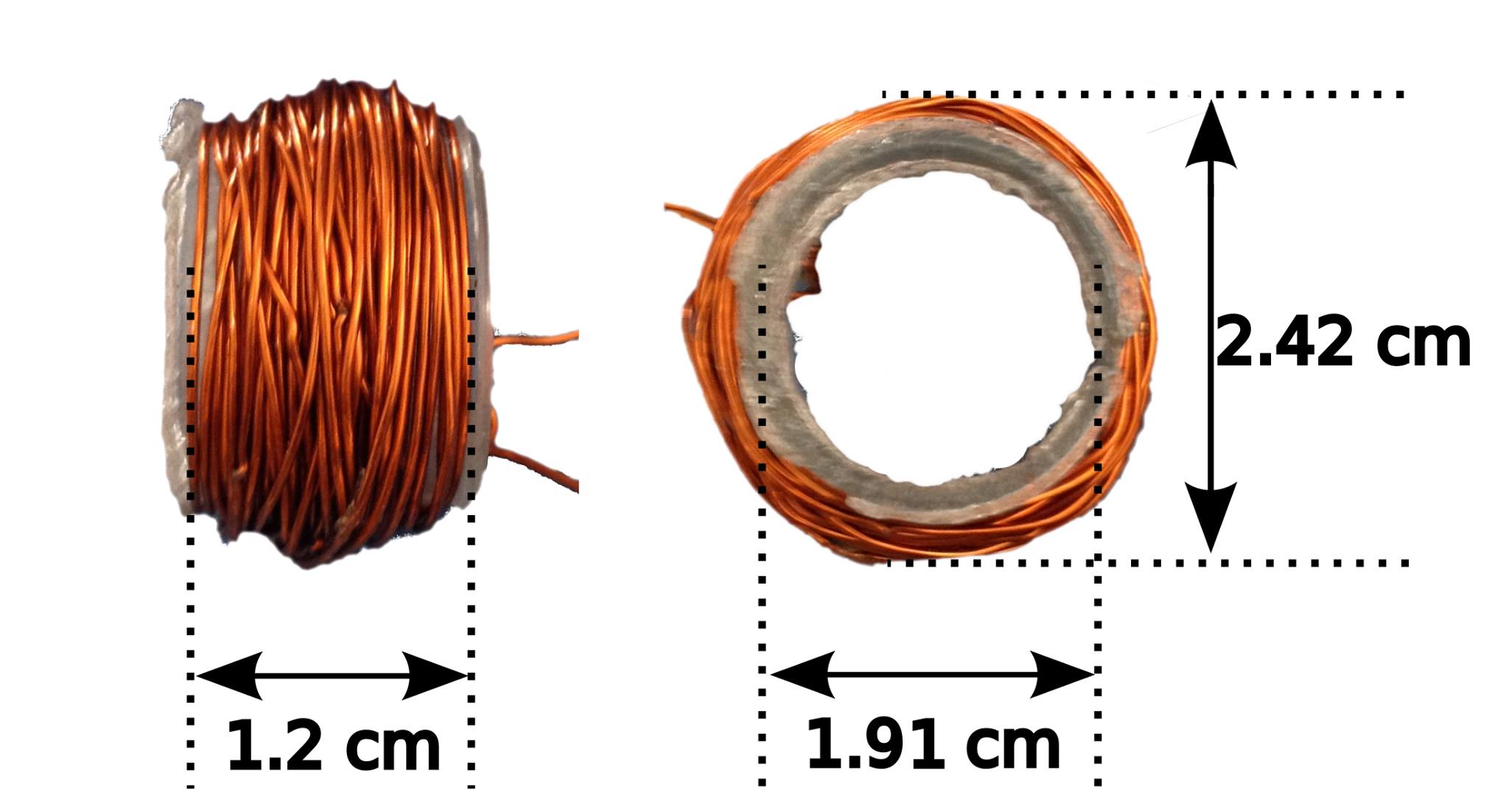}
\caption{(Color online) Spatial dimensions of the coil used in this work. Notice that the coil width is 1.2 cm, we use its half width as the referential point to measure the distance z from the magnetic field sensor. The inner and outer diameter are 1.910 cm  and 2.440 cm, respectively.}
\label{fig:coil}
\end{figure}

We present in figure \ref{fig:coil} the dimensions of the coil used. From figure \ref{fig:coil} one finds $R_{M} = (1.910 + 2.440)/4\ cm=1.088\ cm$, where $R_{M}$ is the mean radius of the coil.
 We also know the electrical current value and the number of turns of the coil. These informations allow one to obtain an estimation for the magnetic permeability $\mu_{air} \cong \mu_{0}$. The inverse cubic dependence of the magnetic field for the coil is consistent with the magnetic field generated by a pure magnetic dipole (m) in its axis\cite{griffiths}, given by

\begin{equation}
\vec{B}(z) = \frac{\mu_{0}}{2\pi}\frac{m}{z^{3}} \hat{z} \,\,\,,
\label{eq:magnetic_dipole}
\end{equation}
\noindent
where $m = NI\pi R_{M}^{2}$, $N = 62$ turns, $I = 0.9\ A$ and $R_{M}$ is the mean radius of our coil.  
Therefore, we impose $b=-3$ for our data and we leave $a^\prime$ as the only parameter in the data fit. The data fit performed for the points displayed in figure \ref{fig:B_X_d_coil} returns $a^\prime = 4240 \pm 25.81$, with a standard deviation less than $1 \%$.  The relation between the $a^\prime$ coefficient for the coil and $\mu_{air}$ is given by  

\begin{equation}	
\mu_{air} \cong \frac{a^\prime 2\pi}{m} = \frac{2a^\prime}{NIR_{M}^{2}} \,\,\,.
\end{equation}

Converting all the units to their SI values, the result leads to $\mu_{air} \cong 1.298 \times 10^{-6} Hm^{-1}$, in good agreement with the expected value of $\mu_{0} \equiv 4\pi \times 10^{-6} \approx 1.2566370614 \times 10^{-6} Hm^{-1}$. 

Although this estimation for $\mu_{air}$ is already good enough, we also perform another procedure to evaluate the air permeability and the
error analysis.

\begin{table}[!htb]
\caption{Magnetic field $B\left(z\right)$ and $z$ for the coil.}
\begin{tabular}{ | c | c | }
\hline
    Magnetic field $(B)$ & Axial distance $(z)$    \\ 
    $[\mu T]$  &  $[cm]$ \\ \hline
    $38.5$  & $4.8 $  \\ \hline
    $29$ & $5.3$   \\ \hline
    $ 21.5$ & $5.8$    \\ \hline
    $16.5$ & $6.3$     \\ \hline
    $13$ & $6.8$    \\ \hline
    $10.5$ & $7.3$     \\ \hline
    $9$ & $7.8$      \\ \hline
    $7.5$ & $8.3$     \\ \hline
    $6.5$ & $8.8$   \\ \hline        
\end{tabular}
\label{table:tabela2}
\end{table}

\begin{figure}[!htb]
\includegraphics[scale=0.68]{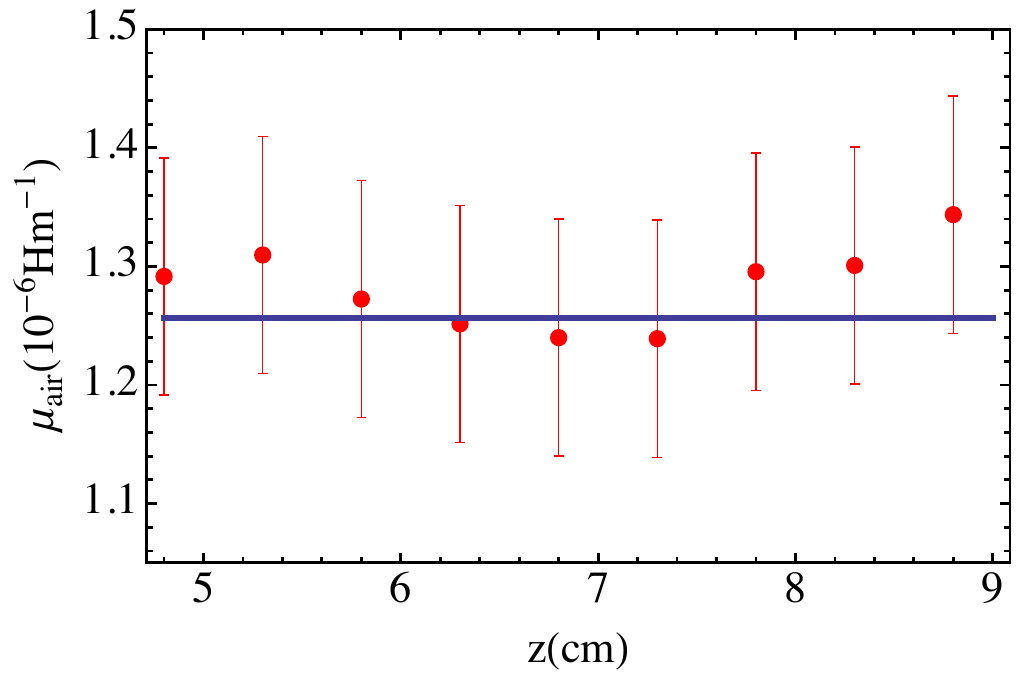}
\caption{(Color online) Air permeability values obtained from the expeimental data evaluated at different axial distances (red points). The blue line indicated the established value for $\mu_{air}$. }
\label{fig6}
\end{figure}

 We use the coil data points from Table \ref{table:tabela2}, and we replace these values in equation \ref{eq:magnetic_dipole} in order to find the value of $\mu_{air}$ for each single point. 
We also consider the following uncertainties in the experimental measurements: $\delta B = 0.5 \mu T$, $\delta I = 0.01\ A$, $\delta z = 0.001\ m$, $\delta N = 1$ and $\delta R_{M} = 0.00005\ m$. There is a precision difference between $R_M$ and $z$  because we used different measurement devices. For $z$ we use a simple ruler, and for $R_{M}$ we use a calliper rule. For the error calculation we use the following variance formula taking into account all the independent variables \cite{error_analysis}:

\begin{equation}
\sigma_{\mu_{air}} = \sqrt{\Big(\frac{\partial \mu_{air}}{\partial B}\Big)^2 \delta B^{2} + \ldots + \Big(\frac{\partial \mu_{air}}{\partial R}\Big)^2 \delta R^{2}} \,\,\,.
\label{eq:error_calculation}
\end{equation}
Each partial derivative of the previous expression is evaluated at the average values of magnetic field and of axial distances, in such a way that we obtain the same uncertainty value for all the points, as a mean value. The uncertainty $\sigma_{\mu_{air}} = 0.1 \times 10^{-6}Hm^{-1}$ tell us that the experiment enables the evaluation of $\mu_{air}$ with 2 significant figures. We show in Fig.\ \ref{fig6} the $\mu_{air}$ values obtained for different axial distances, given by red dots with correspondent uncertainties. The expected value of $\mu_{air}$ is also exhibited in the figure given by the blue line. Notice the relatively small data deviations from the expected value, which means that following this procedure, we also obtained a fair estimate for $\mu_{air}$.

Unfortunately, in this experiment analysis it was not possible to determine the value of $\mu_{air}$ using the magnet. In fact, all that one is able to make is an estimate for the magnet magnetic dipole $m$, assuming the value for $\mu_{air}$.

From the circuit made we obtained a linear dependence between B and I. In addition, we observed the same proportionality $B \propto z^{-3}$ for both the coil and the magnet, enabling one to discuss the parallel between them. Finally, we also could make a fair and simple estimate for the magnetic permeability $\mu_{air}$, even under the limitations of our experimental device.
Unfortunately, in our analysis applied to the magnet does not give the value of $\mu_{air}$. In the magnet case, all we can do, assuming the value for $\mu_{air}$, is an estimate for its magnetic dipole $m$.
For further experiments we suggest the study of the magnet dependence on distance for other geometries, like the long straight wire or the current on a plane sheet of steel.

\section*{Acknowledgements:}

This work was partially supported by CNPq and CAPES (Brazilian Government Agencies).

%
%
%
%
%


\begin{thebibliography}{40}
\bibitem{footnote} To determine the inner distance $d$ we use a needle. As the needle is ferromagnetic, it will influence the magnetometer. Strolling with the needle tip on the screen, the position in which the magnetic field is at his maximum should be the location of the magnetometer inside the tablet or phone.
\end{thebibliography}
\end{document}